\documentclass[12pt]{article}
\usepackage{amssymb,amsmath,epsfig}
\begin{document}
\parindent=0pt
\parskip=6pt
\rm

\begin{center}
{\bf Some remarks on the magnetic phase diagram of ferromagnetic superconductor UGe$_2$}\\
\vspace{0.5cm}
{\bf Diana V. Shopova}\\
{\em Collective Phenomena laboratory, G. Nadjakov Institute
of Solid State Physics,\\
 Bulgarian Academy of Sciences, BG-1784 Sofia, Bulgaria.}
\end{center}

E-mail: sho@issp.bas.bg

\begin{abstract}
The phase transition to ferromagnetic order in itinerant ferromagnetic superconductor UGe$_2$ as function of pressure shows change of phase transition order. At low temperature and high pressure  the transition is of first order and at pressure of about 1.42 GPa the order changes to second. On the basis of Landau expansion of free energy up to the sixth order in magnetization we calculate the phase diagram taking into account the magneto-elastic interaction as the mechanism responsible for this change.\\
We propose a simple Stoner-like dependence of the Curie temperature on the pressure and present the results for measurable thermodynamic quantities as the entropy jump at the first order phase transition.
\end{abstract}

\section {Introduction}
The coexistence of ferromagnetism and bulk superconductivity
in Uranium-based intermetallic compound, UGe$_2$~\cite{Saxena:2000,Huxley:2001} has been a subject of intensive research since its discovery. The experimental results show that the UGe$_2$ is an itinerant  ferromagnet with orthorhombic symmetry  and can be well described as uniaxial ferromagnet of Ising type. The superconducting phase exists entirely within the ferromagnetic domain at low temperature and pressure interval between 1.0 $\div$ 1.5 GPa. At T=0 and ambient pressure UGe$_2$ orders ferromagnetically through second order phase transition with Curie temperature of T$_0$ = 52 K. There are two distinct ferromagnetic phases usually denoted by FM2 and FM1 with different magnitude of magnetic moments. With increasing pressure,the magnetic ground state switches from a highly polarized phase (FM2, M$_0$ $\sim$ 1.5 $\mu$ B) to a weakly polarized phase (FM1, M$_0$ $\sim$ 0.9 $\mu$ B) at pressure of P$_x\sim$ 1.2 GPa.  When the pressure is increased the order of transition from paramagnetic to  FM1  changes from second to first at the tricritical point T$_{cr}$ on the T(P) diagram, and at some critical pressure P$_c$, the ferromagnetic order disappears. Detailed description of P-T phase  diagram of UGe$_2$  can be found in~\cite{Aoki:2011}; schematic illustration is given in Fig.~1.\\
\begin{figure}
\begin{center}
\epsfig{file=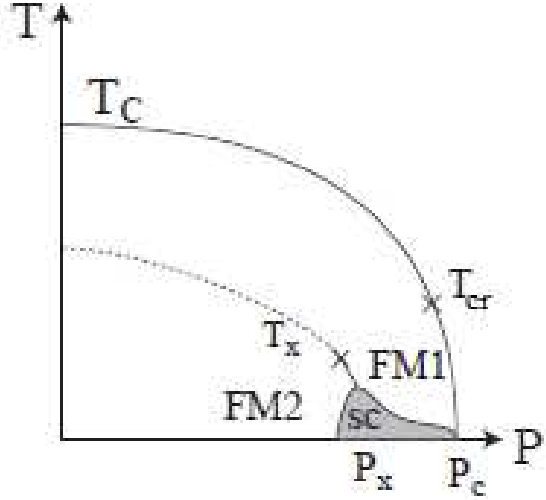,angle=-0, width=8cm}
\end{center}
\caption{\footnotesize An illustration of $T-P$ phase diagram
of UGe$_2$: FM1 - high-pressure ferromagnetic phase, FM2 - low-pressure ferromagnetic phase, SC-phase of coexistence of ferromagnetism and superconductivity.
$T_{x}(P)$ and $T_{c}(P)$ are the
respective magnetic phase transition lines; $T_{cr}$ is tricritical point;
$P_c$ is the critical pressure, at which the ferromagnetic order disappears.} \label{Fig1}
\end{figure}
The steep  drop of T(P) to zero at the critical pressure  P$_c$  in  UGe$_2$ is generally agreed to be an evidence of first order phase transition to the paramagnetic phase, see for example~\cite{Pfleiderer:2002, Pleiderer:2009} and is of great interest as a candidate for quantum critical point ~\cite{Kotegawa:2011}. Recently Mineev~\cite{Mineev:2012} reexamined the development of instability with respect to the continuous type of transition from paramagnetic phase  to FM1 on the basis of magneto-elastic mechanism, treating it in mean-field approach. The influence of pressure on magnetic fluctuations close to the critical point leads only to weak first-order transition at P$_c$ in contradiction to the experimental results,~\cite{Pleiderer:2009}. Mineev made an estimate of the theoretical results with the experimentally obtained parameters for UGe$_2$  and concluded that most likely the order of the phase transition is changed from second to first because the coefficient before the fourth order term in Landau expansion of the free energy may become negative due to interaction of magnetic and elastic degrees of freedom.\\
In this paper we shall study within the phenomenological Landau energy the effect of pressure and compressibility on the phase transition order from paramagnetic phase to FM1 in UGe$_2$ by including a sixth order term in the Landau expansion of the magnetic free energy and considering the possible change with the pressure of the sign of parameter at the forth order in magnetization.
\section {Landau free energy}
In previous papers~\cite{Cottam:2008, Shopova:2009} we derived and applied Ginzburg-Landau free energy of UGe$_2$ in the following form:
\begin{equation}
\label{Eq1} f(\vec{M},\vec{\psi}) = f_M + f_{sc} + f_{int},
\end{equation}
\noindent where, f$_{sc}$ is the free energy density describing pure superconducting system:
\begin{equation}
\label{Eq2}f_{sc}=a_s |\mathbf{\psi}|^2 + \frac{b_s}{2} |\mathbf{\psi}|^4 +\frac{u_s}{2} |\mathbf{\psi}^2|^2 + \frac{v_s}{2}\sum_{j=1}^3{|\psi_j|^4},
\end{equation}
with $\mathbf{\psi}$ the superconducting order parameter, $a_s=\alpha_s(T-T_s)$ and $b_s >0$ are the standard Landau parameters; the terms with $u_s$ and $v_s$ represent the anisotropy of spin triplet Cooper pairs and crystal anisotropy, respectively.
 The ferromagnetic energy density up to the forth order in magnetization $\mathbf{M}$ is denoted by f$_M$
 \begin{equation}\label{Eq3}
 f_M= a_f |\mathbf{M}|^2 + \frac{b_f}{2} |\mathbf{M}|^4,
  \end{equation}
  where $a_f=\alpha_f(T-T_f(P))$, and $b>0$.
f$_{int}$ stands for the interaction between superconducting and magnetic order parameters:
  \begin{equation}\label{Eq4}
f_{int} = i\gamma_0 \mathbf{M}.(\mathbf{\psi}\times \mathbf{\psi}^\ast) + \delta \mathbf{M}^2|\mathbf{\psi}|^2
  \end{equation}
The pressure effect in (\ref{Eq1}) was included phenomenologically only through the linear dependence of temperature of ferromagnetic transition $T_f$ on pressure. This simple free energy gives surprisingly good general results in description of UGe$_2$ phase diagram and its comparison to the experimental data. The temperature of transition to the phase of coexistence of superconductivity and ferromagnetism in UGe$_2$ is always lower than the Curie temperature, a fact that confirms the idea of superconductivity that is triggered by the spontaneous magnetization \textbf{M}. That is why the detailed description of the purely magnetic phase transitions is very important for the construction of the total phase diagram.
The most substantial deviations from the experimental $T(P)$ for UGe$_2$  given by the free energy (\ref{Eq1}) are two: the first is concerned with the overestimation of the critical pressure where the superconductivity appears for the first time when lowering the temperature, which is represented by a maximum on the phase diagram - see Fig.1. This is likely due to the fact that magnetic free energy,~(\ref{Eq3}) does not properly describe the whole magnetic phase diagram of  UGe$_2$, especially the presence of two magnetic phases, as the phase transition between the two magnetic phases is at pressure,  very close to the  pressure of superconducting transition. This problem is widely discussed in the literature and needs additional studies.
Another feature that is lacking in the description by free energy~(\ref{Eq1}) is the change of order from second to first of the transition from paramagnetic to ferromagnetic ordering of FM1. \\
Taking into account the arguments of Mineev~\cite{Mineev:2012} about the instability of second order transition to ferromagnetic ordering in UGe$_2$  due to the strong dependence of Curie temperature on pressure which may lead  to change of the sign of $b_f$ in~(\ref{Eq3}) from to positive to negative, we will consider in detail this effect. Moreover, the neutron scattering experiments of lattice expansion~\cite{Sokolov:2011} show that the magnetoelastic coupling in UGe$_2$ is strong with substantial influence on the phase transitions.\\
We expand the  magnetic part $f_M$ of Ginzburg-Landau energy  to sixth order in magnetization $\mathbf{M}$ and take into account the lowest order of interaction between the magnetic and elastic degrees of freedom. As UGe$_2$ is highly anisotropic uniaxial ferromagnet  without loss of generality we may write that  $\mathbf{M} = (0,0,M)$. In the absence of shear deformation magnetic part $f_M$ of free energy will become:
\begin{equation}
 \label{Eq5} f_M (M, \varepsilon) = a(T,P)M^2+\frac{b}{2}M^4+\frac{c}{3} M^6+\frac{K}{2}\varepsilon^2-q\varepsilon M^2
\end{equation}
The variable $\varepsilon$  is the relative volume change, M is the magnetization, $K > 0$ is the bulk modulus. We assume that the parameters $b$ and $c$ do not depend on temperature and pressure, so $b \equiv b_0 >0$ and $c \equiv c_0 >0$, where $b_0=b(P=0,T=0)$ and $c_0=c(P=0,T=0)$. The dependence of $a(P,T)$ is given by the relation $a(P,T)=\alpha_0 [T-T_f(P)]$. The term $q\varepsilon M^2$ is the coupling energy between magnetization and the relative volume change with coupling parameter $q$.
Similar form of the magnetic free energy is proposed in~\cite{Yang:2008}, where a first order phase transition due to effect of compressibility is discussed for transition metals Ni, Fe, Co.\\
We minimize the energy~(\ref{Eq5}) with respect to relative volume change $\varepsilon$ in absence of shear deformation, $\partial{f_M}/\partial{\varepsilon} =0$ and obtain:
\begin{equation}\label{Eq6}
f_M (M) = a(T,P)M^2+\frac{1}{2}\left(b_0 -\frac{q^2}{K}\right) M^4+\frac{c_0}{3} M^6.
\end{equation}
Thus the coupling of magnetic and elastic degrees of freedom modifies the coefficient  in front of the fourth order term in~(\ref{Eq6}) and it may become negative if:
$$\frac{q^2}{K b_0} >1.$$
The coupling parameter $q$ is the slope of $T_f$ volume dependence~\cite{Bean:1962}, or $q = \alpha_0 (dT/d\varepsilon)$, and using the relation $\varepsilon= - K P$, it can be written in the form:
\begin{equation}\label{Eq7}
    q = -\alpha_0 K \frac{dT_f}{dP}
\end{equation}
Therefore, in certain range  of temperatures and pressures the ferromagnetic ordering may occur through first order phase transition from the disordered phase. This depends crucially on the steepness of the function $T_f(P)$, which is measured by the derivative of $T_f(P)$ with respect to pressure. The experiments~\cite{Aso:2006} show, for example, that the  Stoner-Wohlfarth theory for weak itinerant ferromagnets  can fairly well describe the behavior of magnetization with pressure in low-pressure ferromagnetic phase, denoted in Fig.~1 as FM2. There the pressure derivative will take the form:
\begin{equation}\label{Eq8}
\frac{dT_f}{dP}=-\frac{\gamma}{T_f},
\end{equation}
 and $\gamma$ is connected to the microscopic model used for calculation of~Eq.(\ref{Eq8}). From the above relation it is straightforwardly seen that the dependence of $T_f$ on pressure will be proportional to $\sqrt{1-P/P_c}$, where $P_c$ is the critical pressure of disappearance of ferromagnetism.
This Stoner dependence of $T_f$ on pressure renormalizes the parameter $b=(b_0 -q^2/K)$ in front of the forth order term in magnetization in~(\ref{Eq6}) in the following way:
\begin{equation}\label{Eq9}
b =b_0\left[1-\nu_0 \left(1-\frac{P}{P_c}\right)^{-1}\right],
\end{equation}
where $\nu_0=\alpha_0^2K \gamma^2/(b_0T_0^2)$.
It is obvious that $b$ can become negative which leads to instability of second order phase transition, but the above relation can hardly be  applied to the description of the first order phase transition from paramagnetic to FM1 phase as close to $P_c$, the parameter $b$ from~(\ref{Eq9}) becomes divergent and grows in unlimited way. So Stoner-Wohlfarth theory can hardly be directly applied to  the description of the first order phase transition in its original form.
Here we will use the interpolation approach, proposed in~\cite{Mohn:1988} for the dependence of Curie temperature of the second-order phase transition on pressure, which is a solution of the following equation:
\begin{equation}\label{Eq10}
\tau^2 t^2 +(1-\tau^2) t +p-1=0.
\end{equation}
In the above equation $t=T_f(P)/T_0$ and $T_0=T_f(P=0)$; the dimensionless variable, $p$, is defined by $p=P/P_c$ with $P_c$ the critical pressure, at which the ferromagnetic ordering disappears. The quantity $0\leq\tau\leq 1$ is the measure that accounts for the spin fluctuations with respect to the single particle Stoner excitations. The limit $\tau=1$ gives the Stoner dependence of the critical temperature on pressure, and $\tau=0$ means that the spin fluctuations determine $T_f(P)$.\\
\section{Results and discussion}
The experiments show that for $T=0$ and $P=0$ the phase transition from paramagnetic phase to FM1 phase is of second order, and using that, we can write the free energy~(\ref{Eq6}) in dimensionless form:
\begin{equation}\label{Eq11}
f\equiv \frac{F}{cM_0^6} = \mu (t-t_f)m^2+\mu \frac{b}{2} m^4+\frac{1}{3}m^6;
\end{equation}
where, $M_0=\sqrt{\alpha_0T_0/b_0}$ is the magnetization at $T=0$, $P=0$; $t=T/T_0$ and $\mu=b_0^2/(\alpha_0 cT_0)$. The dimensionless critical temperature, $t_f$, obtained from equation~(\ref{Eq10}) will be:
\begin{equation}\label{Eq12}
t_f=\frac{T_f(p)}{T_0}=\frac{1}{2\tau^2}\left[\sqrt{(1+\tau^2)^2-4\tau^2 p}-(1-\tau^2)\right].
\end{equation}
The parameter $b$ in front of the forth order term of the free energy~(\ref{Eq11}) becomes:
\begin{equation}\label{Eq13}
b=1-\frac{\nu}{(1+\tau^2)^2-4\tau^2 p},
\end{equation}
where $\nu=(T_0K/P_c^2) \Delta C$ and $\Delta C$ is the specific heat jump of the second order phase transition at $P=0$. For pressures in the interval $[(1-\tau^2)^2-\nu]/(4\tau^2)\leq p < (1-\tau^2)^2/(4\tau^2)$, the parameter $b<0$. The equality is fulfilled for pressure $p_{cr}$ of the tricritical point and there $b=0$. For UGe$_2$ the tricritical pressure $p_{cr}\sim 0.94\div 0.95$ according to the different experimental data, namely $P_{cr} \sim 1.41\div 1.42$ GPa and $P_c \sim 1.5$ GPa.\\
The equation of state for free energy~(\ref{Eq11}) will be:
\begin{equation}\label{Eq14}
\mu (t-t_f)m +\mu b m^3+m^5=0
\end{equation}
with one stable solution:
\begin{equation}\label{Eq15}
m^2=\frac{1}{2}\left[-\mu b+\sqrt{\mu(\mu b^2-4 (t-t_f))}\right]
\end{equation}
existing for $t\leq \mu b^2/4+t_f$. It is seen from the above relation that for $b < 0$, the magnetization $m$ will be positive at $t=0$ and its jump at $P=P_c$ will be:
$$\frac{\mu}{2} \left[\frac{\nu}{(1-\tau^2)^2}-1 +\sqrt{(1-\frac{\nu}{(1-\tau^2)^2})^2+\frac{4}{\mu}}\right]$$
The equilibrium temperature for the first order phase transition is determined in the standard way by finding the minimum of free energy~(\ref{Eq11}) or $f(m)=0$, with $m$ given by Eq.~(\ref{Eq15}). With the help of equation of state, (\ref{Eq14}) we obtain for the equilibrium energy:
\begin{equation}\label{Eq16}
f_{eq}=\frac{\mu}{12} m^2 \left[8 (t-t_f)-b^2 \mu +b \sqrt{\mu(\mu b^2-4 t+4 t_f)}\right]
\end{equation}
For
\begin{equation}\label{Eq17}
t_{eq}=t_f+\frac{3 \mu b^2}{16}
\end{equation}
$f_{eq}=0$ and paramagnetic and FM1 phase coexist. In the interval of temperatures $t_f<t<t_{eq}$, the FM1 phase is stable.
For UGe$_2$ the calculations with the experimental data available, give that $\tau\sim 0.8$ which according to~\cite{Mohn:1988} means that the Stoner mechanism is prevailing with corrections from spin fluctuations. The numerical estimate of $t_{eq}$ with this value of $\tau$ gives that the equilibrium temperature of first order phase transition is $\sim 11.5$ K at the measured critical pressure of $P_c=1.5$ GPa. The dimensionless latent heat $Q$, at the first-order transition temperature $t_{eq}$, see Eq.~(\ref{Eq17}), is given by $Q=-1/4\mu^2 b (3/8 \mu b^2+t_f)$. At the critical pressure $p_c= 1$, the numerical value of $Q$ is $\sim$ 0.11.\\
The crucial point in the description of the first-order phase transition from paramagnetic to FM1 phase is the explicit dependence of Curie temperature on pressure. Recently in~\cite{Konno:2012}, the pressure coefficient of Curie temperature is calculated from microscopic model numerically. The results show that $dT_f/dP$ depends in a complex way on two terms coming from the derivatives of Fermi temperature and exchange integral with  pressure, respectively. But these results cannot be applied directly in phenomenological analysis. More theoretical calculations are needed in this respect as the first order of the phase transition is also important for the understanding of phase of coexistence of superconductivity and ferromagnetism.
\section{Conclusion}
The presented phenomenological analysis shows that the account of  compressibility to the lowest order in the relative volume change can explain the first order phase transition between FM1 and paramagnetic phase in UGe$_2$. The use of phenomenological interpolation relation derived in~\cite{Mohn:1988} adequately describes qualitatively the first order phase transition. In order to properly understand the nature of the this transition, more precise microscopic calculation are needed for the dependence of Curie temperature on pressure in the high-pressure ferromagnetic phase of UGe$_2$.

\end{document}